\documentclass[aps,prl,twocolumn,superscriptaddress]{revtex4}
\usepackage{float}
\usepackage{makeidx}

\usepackage{graphicx}
\usepackage{amsmath}

\begin{document}

\title{Observation of the Efimovian Expansion in Scale Invariant Fermi Gases}

\author{Shujin Deng}
\thanks{They contribute equally to this work}
\affiliation{State Key Laboratory of Precision Spectroscopy, East China Normal University, Shanghai 200062, P. R. China}

\author{Zhe-Yu Shi}
\thanks{They contribute equally to this work}
\affiliation{Institute for Advanced Study, Tsinghua University, Beijing, 100084, P. R. China}

\author{Pengpeng Diao}
\affiliation{State Key Laboratory of Precision Spectroscopy, East China Normal University, Shanghai 200062, P. R. China}

\author{Qianli Yu}
\affiliation{State Key Laboratory of Precision Spectroscopy, East China Normal University, Shanghai 200062, P. R. China}

\author{Hui Zhai}
\affiliation{Institute for Advanced Study, Tsinghua University, Beijing, 100084, P. R. China}

\author{Ran Qi}
\email{qiran@ruc.edu.cn}
\affiliation{Department of Physics, Renmin University of China, Beijing, 100872, P. R. China}

\author{Haibin Wu}
\email{hbwu@phy.ecnu.edu.cn}
\affiliation{State Key Laboratory of Precision Spectroscopy, East China Normal University, Shanghai 200062, P. R. China}

\begin{abstract}

Scale invariance emerges and plays an important role in strongly correlated many-body systems such as critical regimes nearby phase transitions and the unitary Fermi gases. Discrete scaling symmetry also manifests itself in quantum few-body systems such as the Efimov effect. Here we report both theoretical predication and experimental observation of a novel type expansion dynamics for scale invariant quantum gases. When the frequency of the harmonic trap holding the gas decreases continuously as the inverse of time $t$, surprisingly, the expansion of cloud size exhibits a sequence of plateaus. Remarkably, the locations of these plateaus obey a discrete geometric scaling law with a controllable scale factor and the entire expansion dynamics is governed by a log-periodic function. This striking expansion of  quantum Fermi gases shares similar scaling laws and same  mathematical description as the Efimov effect.  Our work demonstrates the first expansion dynamics of a quantum many-body system with the temporal discrete scaling symmetry, which reveals the underlying spatial continuous scaling symmetry of the many-body system.

\end{abstract}

\maketitle

Interaction between dilute ultracold atoms is described by the $s$-wave scattering length. For a spin-$1/2$ Fermi gas, when the scattering length diverges at a Feshbach resonance, there is no length scale other than the interparticle spacing in this many-body system, and therefore the system, known as the unitary Fermi gas, becomes scale invariant. The spatial scale invariance leads to universal thermodynamics and transport properties as revealed by many experiments \cite{thermodynamics0,thermodynamics1,thermodynamics2,thermodynamics3,thermodynamics4,thermodynamics5,transport0,transport1,transport2,transport3,transport4,transport5,scale}. On the other hand, in a boson system with an infinite scattering length, three-body bound state can form, where an extra length scale of the three-body parameter sets a short-range boundary condition for all three bosons being very close. It turns the continuous scaling symmetry into a discrete scaling symmetry, and gives rise to infinite number of three-body bound states whose energies obey a geometric scaling symmetry. This is well known as the Efimov effect \cite{Efimovtheory1,Efimovtheory2}, which has been observed in quite a few cold atom experiments \cite{Efimov_exp1,Efimov_exp2,Efimov_exp3,Efimov_exp4,Efimov_exp5,Efimov_exp6,Efimov_exp7,Efimov_exp8,Efimov_exp9}, and recent experiments have also confirmed the geometric scaling of the energy spectrum \cite{scaling_exp1,scaling_exp2,scaling_exp3,scaling_exp4}. Both the continuous and the discrete scaling symmetry are interesting emergent phenomena in a strongly interacting system.

\begin{figure}[tp]
\includegraphics[width=3.2 in]
{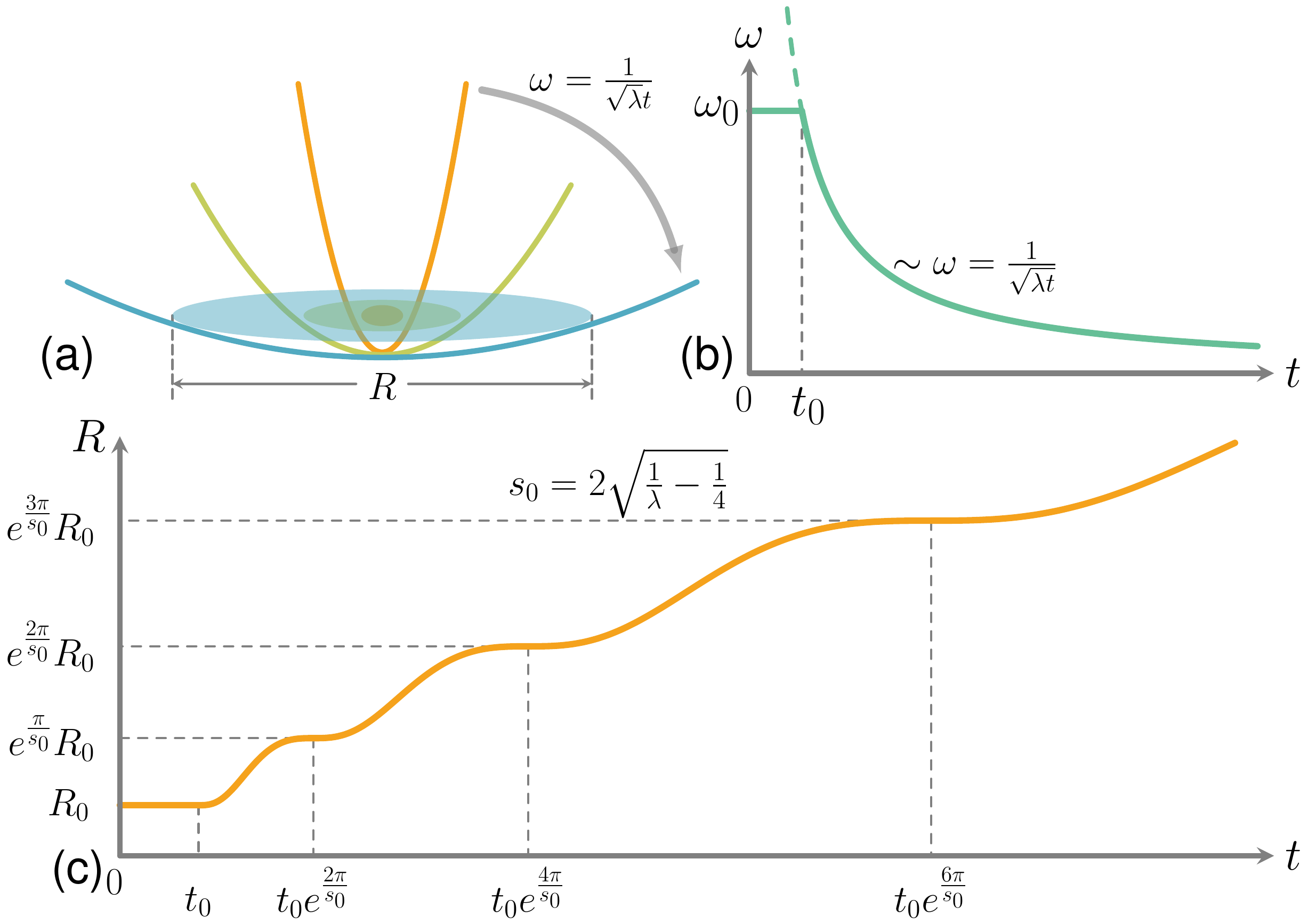}
\caption{ (a-b) The schematic of the Efimovian expansion: A scale invariant ultracold gas is first hold in a harmonic trap with frequency $\omega_0$. Then, starting from $t_0=1/(\sqrt{\lambda}\omega_0)$, the trap frequency starts to decrease as $1/(\sqrt{\lambda}t)$, and the cloud expands. (c) The theoretical predication of the Efimovian expansion: the cloud size $\mathcal{R}$ as a function of time $t$ follows a log-periodic function and exhibits a series of plateaus. The locations of the plateaus obey a geometric scaling law.  }
\label{illustration}
\end{figure}

For a harmonic trapped gas, the expansion dynamics offers great insight to the property of the gas. Well known example is the anisotropic expansion that proves hydrodynamics behavior due to the Bose condensation \cite{BEC1,BEC2} or strong interactions of Fermi gas \cite{Thomas}. Other examples are, for instance, slowing down of expansion in a disorder potential provides evidence for localization behaviors \cite{localizations1,localizations2} and expansion in the presence of optical lattice reveals correlation effects \cite{Bloch}. In this work, we ask a question that, considering a scale invariant quantum gas hold by a harmonic trap, when the trap is gradually opened up with decreasing the trap frequency $\omega$ as $1/(\sqrt{\lambda} t)$ ($\lambda$ is a coefficient and $t$ is the time), as shown in Fig. \ref{illustration}(a) and (b), how does the gas expand? Naively, by dimensional analysis, one would expect that the cloud size $\mathcal{R}$ just increases as $\sqrt{t}$. Here we show, both theoretically and experimentally, that it is not the case. When $\lambda$ is smaller than a critical value, the expansion dynamics displays a discrete scaling symmetry in the time domain. As shown in Fig. \ref{illustration}(c), $\mathcal{R}$ as a function of $t$ displays a sequence of plateaus, which mean that at a set of discrete times $t_n$ the cloud expansion surprisingly stops, despite of the continuous decreasing of the trap frequency. The locations of the plateaus $t_n$ obey a geometric scaling behavior. More interestingly, this striking behavior of a discrete scaling law in the time domain is in fact a consequence of the continuous spatial scaling symmetry.

To explain this intriguing dynamics, we shall first point out the insight that why $\omega=1/(\sqrt{\lambda} t)$ is so special. For simplicity, we first consider a three-dimensional isotropic trap $V(r)=m\omega^2 R^2/2$. Let us consider that a many-body system, if in the absence of a trapping potential, is invariant under a scale transformation ${\bf r}\rightarrow \Lambda {\bf r}$, while in the presence of a static harmonic trap, the fixed harmonic length introduces an additional length scale that could break this spatial scale invariance. Nevertheless, if $\omega$ changes as $1/(\sqrt{\lambda} t)$, the time-dependent Schr\"odinger equation exhibits a space-time scaling symmetry under the transformation ${\bf r}\rightarrow \Lambda {\bf r}$ and $t\rightarrow \Lambda^2 t$. With the scaling symmetry, it is straightforward to derive that the equation-of-motion for the cloud size $\langle\hat{R}^2\rangle$ is given by (see supplementary material for the detailed derivation):
\begin{eqnarray}
\frac{d^3}{dt^3}\langle \hat{R}^2\rangle+\frac{4}{\lambda t^2}\frac{d}{dt}\langle \hat{R}^2\rangle-\frac{4}{\lambda t^3}\langle \hat{R}^2\rangle=0.\label{Rt}
\end{eqnarray}
Obviously, the differential equation is invariant under a continuous scaling of time $t$. However, in practices, one should always start with a finite initial trap frequency $\omega_0$ before turning it down, which corresponds to an initial time $t_0$ with $\omega_0=1/(\sqrt{\lambda} t_0)$, as shown in Fig. \ref{illustration}(b). The system is at equilibrium for $t<t_0$, and at $t=t_0^+$, $\langle \hat{R}^2\rangle(t_0)=R_0^2$ and $\frac{d^m}{dt^m}\langle \hat{R}^2\rangle|_{t=t_0}=0$ for $m=1,2$. This sets a boundary condition for Eq. \ref{Rt} which can turn the continuous scaling symmetry in the time domain into a discrete one. The solution of this differential equation depends on the value of $\lambda$. When $0<\lambda<4$, the solution is log-periodic function as
\begin{eqnarray}
\frac{\langle \hat{R}^2\rangle(t)}{R_0^2}=\frac{t}{t_0}\frac{1}{\sin^2{\varphi}}\bigg{[}{1-\cos{\varphi}\cdot\cos\bigg(s_0\ln\frac{t}{t_0}+\varphi\bigg)}\bigg{]},\label{Rt_solution}
\end{eqnarray}
where $s_0=2\sqrt{1/\lambda-1/4}$ and $\varphi=-\arctan{s_0}$. Eq. \ref{Rt_solution} clearly reveals the discrete scaling symmetry, i.e. when $t_2=e^{2\pi/s_0}t_1$, $\langle \hat{R}^2\rangle(t_2)=e^{2\pi/s_0}\langle \hat{R}^2\rangle(t_1)$ and $\frac{d^m}{dt^m}\langle \hat{R}^2\rangle|_{t=t_2}=e^{-2\pi(m-1)/s_0}\frac{d^m}{dt^m}\langle \hat{R}^2\rangle|_{t=t_1}$ for all the $m$-th order derivatives. Therefore, at time $t_n=e^{2\pi n/s_0}t_0$, the first- and the second-order time derivatives for $\langle \hat{R}^2\rangle$ become zero and the cloud expansion is strongly suppressed, that is to say, the expansion dynamics shows a series of plateaus around each $t_n$. While when $\lambda>4$, $\langle \hat{R}^2\rangle$ simply follows a power law as $\langle \hat{R}^2\rangle(t)\sim t^{1+\eta}$ for $t\gg t_0$, where $\eta=\sqrt{1-4/\lambda}$.

\begin{figure}[tp]
\includegraphics[width=3.0 in]
{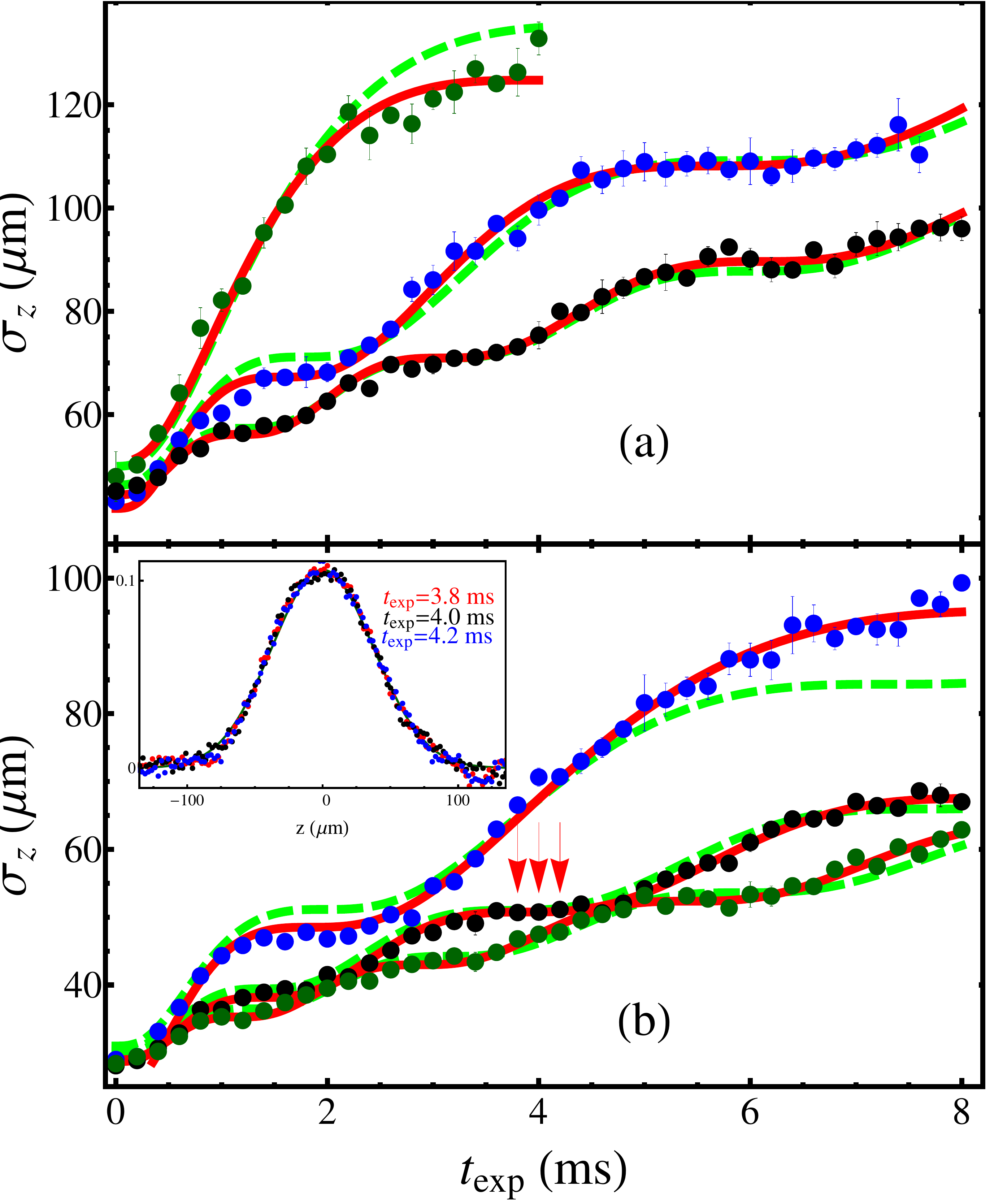}
\caption{$\sigma_z$  (with $\sigma^2_z=2\langle \hat{R}^2_z\rangle$) versus the expansion time $t_\text{exp}=t-t_0$ for a non-interacting Fermi gas of ${}^6$Li measured at $B=528$ Gauss (a) and a unitary Fermi gas measured at $B=832$ Gauss (b). Dots are measured data. Black, blue and green dots denote $\lambda_z=0.02$, $0.07$ and $0.36$ for (a), and $\lambda_z=0.01$, $0.02$ and $0.06$ for (b). The dashed lines are the theory curves based on Eq. \ref{Rt_solution} (with $s_0$ given by Eq. \ref{s0}) without any free parameters, and the solid lines are the best fit using the function form of Eq. \ref{Rt_solution} with $s_0$ as a fitting parameter. The inset in (b) shows three successive density profiles (after the time-of-flight) when the time $t_{exp}$ locates inside a plateau as indicated by the arrows. Error bars represent the standard deviation of the statistic.      }
\label{expansion}
\end{figure}

Here we shall emphasize that this intriguing expansion dynamics is a \textit{universal} phenomenon for all scale invariant quantum gases, which include, for examples, non-interacting gas, the unitary Fermi gas in three-dimension, weakly-interacting gases in two-dimension (when anomaly can be ignored), and the Tonks gas in one-dimension. It is also independent of temperature.

Before proceeding to experimental observation of such dynamical expansion in the ultracold Fermi gases, we would also like to bring out the analogy to the Efimov effect. First, when solving the three-body problem of bosons in the hyper-spherical coordinate, one finally reaches an effective potential as $-1/\rho^2$ ($\rho$ is the so-called hyper-radius) that scales the same way as the kinetic energy, and the Schr\"odinger equation then reduces to a one-dimensional scale invariant differential equation \cite{Efimovtheory1,Efimovtheory2}. Similarly, our equation-of-motion for $\langle \hat{R}^2\rangle$ is also a scale invariant differential equation. Secondly, the short-range boundary condition (i.e. the three-body parameter) plays the similar role as the initial trap frequency here, which sets a boundary condition for the differential equation and turns the symmetry into a discrete scaling symmetry. Thirdly, the solution for the three-body wave function is also a log-periodic function, which is the same function as our solution Eq. \ref{Rt_solution}. Finally, in our case $\lambda$ plays the similar role as the mass ratio in the Efimov problem that controls the scaling factor, as well as whether the effect will occur. Hence, this particular dynamical expansion of a scale invariant gas shares the same symmetry property and similar mathematical description as the three-boson problem of bosons with infinite scattering length. It can be viewed as the counterpart of the Efimov effect in the time domain, and thus is named as ``the Efimovian expansion" here.

In our experiment, we use a balanced mixture of $^6$Li fermions in the lowest two hyperfine states $|\uparrow\rangle\equiv|F=1/2, M_F=-1/2\rangle$ and $|\downarrow\rangle\equiv|F=1/2, M_F= 1/2\rangle$. Fermionic atoms are loaded into a cross-dipole trap to perform evaporative cooling. The resulting potential has a cylindrical symmetry around the propagation axis of the laser and the trap anisotropic frequency ratio $\omega_r/\omega_z$ is about $9$. The trap anisotropy causes an additional complication compared with the isotropic case discussed above. Nevertheless, as shown in the supplementary material, similar results can be obtained for the anisotropic case. Starting at the initial time $t_0$, the trap potential is lowered as
\begin{equation}
V({\bf r})=\frac{m}{2 \lambda_r t^2}r^2+\frac{m}{2\lambda_z t^2}z^2.
\end{equation}
Since $\lambda_r/\lambda_z=(\omega_z/\omega_r)^2 \ll 1$ and the effect is more pronounced along the axial direction than in the transverse direction. (See supplementary material for detailed reason.) Therefore, hereafter we focus on the cloud expansion along the axial direction. Theory shows that the axial cloud square size $\mathcal{R}^2_z$ obeys the same form as Eq. \ref{Rt_solution} except
\begin{equation}
s_0=\omega_\text{b}\sqrt{1/\lambda_z-1/4}, \label{s0}
\end{equation}
where $\omega_\text{b}$ is a factor related to the breathing mode frequency, and $\omega_\text{b}=2$ for the non-interacting gas and $\omega_\text{b}=\sqrt{12/5}$ for the unitary Fermi gas along the axial direction. Feshbach resonance is used to tune the interaction of the atoms either to the non-interacting regime with the magnetic field $B=528$G or to the unitary regime with $B=832$G. The trap frequency is lowered by decreasing the laser intensity, and $\lambda_z$ is controlled by the speed of how fast the laser intensity decreases, with the initial axial trap depth always fixed at $5\%U_0$ where $U_0$ is the full trap potential. Thus, different $\lambda_z$ corresponds to different $t_0=1/(\sqrt{\lambda_z}\omega^0_z)$, where $\omega^0_z$ is the initial axial trap frequency. Finally, after certain expansion time $t_{exp}$ with the trap, the trap is completely turned off and the cloud is probed by standard resonant absorption imaging techniques after a time-of-flight expansion time $t_\text{tof}=200\,\mu s$. Each data point is an average of $5$ shots of the measurements at identical parameters.

\begin{figure}[t]
\includegraphics[width=3.2 in]
{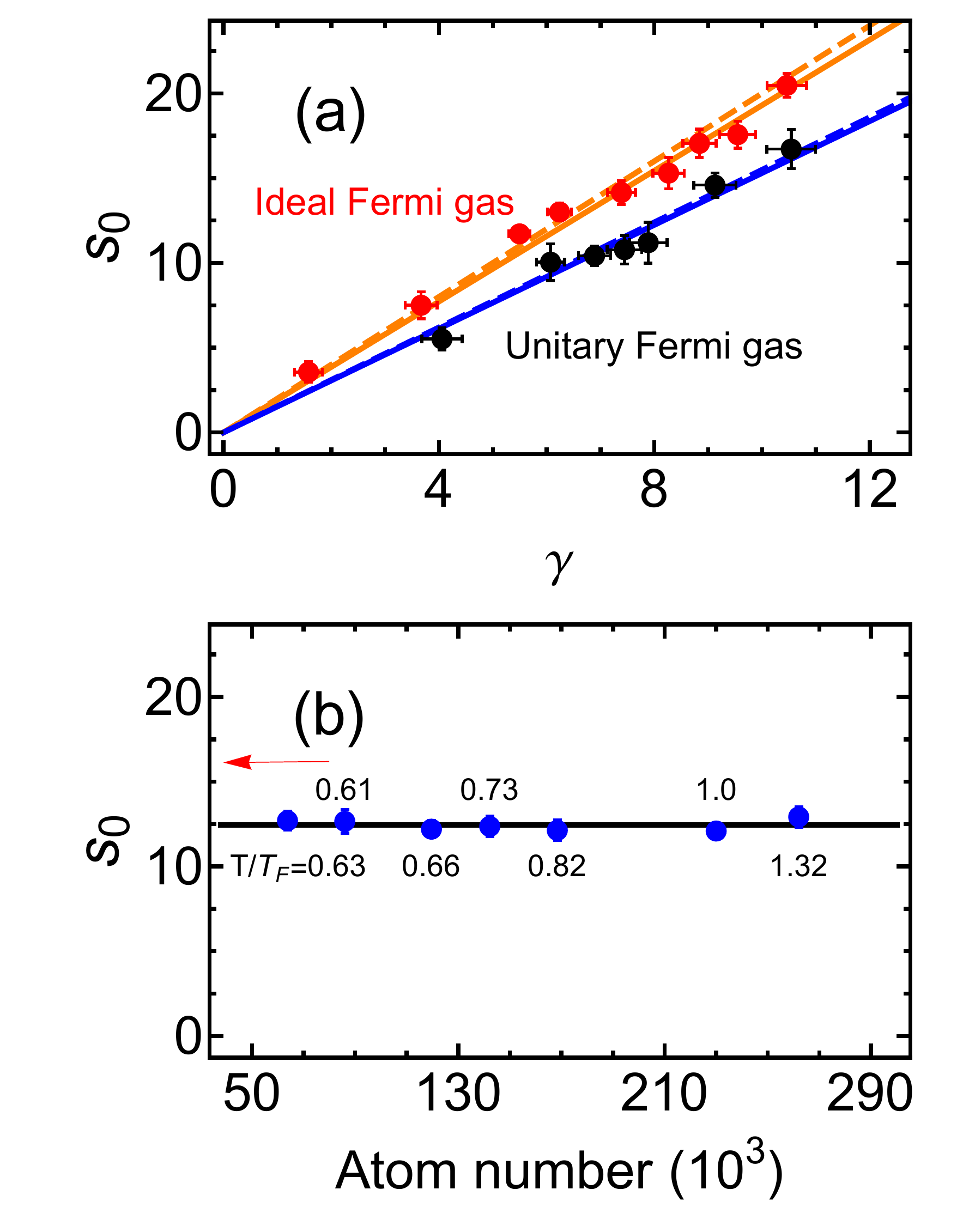}
\caption{(a) $s_0$ obtained from fitting the expansion curves v.s. $\gamma\equiv\sqrt{1/\lambda_z-1/4}$. The solid lines are the linear fitting curves and the dashed lines are $s_0=\omega_\text{b}\gamma$ with $\omega_\text{b}=2$ for the non-interacting fermions and $\omega_\text{b}=\sqrt{12/5}$ for the unitary Fermi gas. (b) For a given $\lambda$ and for the unitary Fermi gas, $s_0$ obtained from fitting the expansion curves for different fermion numbers and temperatures. Solid line is the theory value for the unitary Fermi gas and the arrow indicates the theory value for the non-interacting Fermi gas with same $\lambda$. Error bars in the vertical direction represent the fitting error and the standard deviation of the statistic. Error bars in the horizontal direction represents the standard deviation of the statistic in determining $\lambda$ in repeated measurements. }
\label{s}
\end{figure}

The time-of-flight density profile along the axial direction is fitted by a Gaussian function as $A_0+A_1 e^{-z^2/\sigma^2_{z}}$, from which we obtain $\sigma_{z,\text{obs}}$. $\sigma_{z,\text{obs}}$ is related to the in-situ cloud size by a scale factor $b_z$ via $\sigma_{z,\text{obs}}=b_{z}(t_{\text{tof}})\sigma_z$. $b_z(t_{\text{tof}})$ can be obtained by either hydrodynamic or ballistic expansion equation with the time-of-flight time $t_{\text{tof}}$ (see the supplementary material for detail), with which we can deduce $\sigma_z$ from $\sigma_{z,\text{obs}}$. Since the trap is quite anisotropic, the cloud size expands slowly along the axial direction during a short time-of-flight, and the expansion factor $b_z$ only gives a quantitive correction to the results.  Fig. \ref{expansion} shows the typical measurements of $\sigma_z$ with different $\lambda_z$ for both the non-interacting and the unitary Fermi gases. For instance, for $\lambda_z=0.06$, we decrease the trap frequency from $2\pi \times 567.3\,$Hz to $2\pi \times 71.0\,$Hz within $8\,$ms. Dots are the measured data, and the solid and the dashed lines are both theoretical curves based on Eq. \ref{Rt_solution}, taking $s_0$ as fitting parameter or using $s_0$ given by Eq. \ref{s0}, respectively.  Since $\sigma_{z}$ is obtained by a Gaussian fit to the density profile, $\sigma_z^2=2\langle \hat{R}_z^2\rangle$ and thus the theoretical expression for $\sigma_z/\sigma_{z,0}$ is simply a square root of  Eq. \ref{Rt_solution}. Fig. \ref{expansion} clearly shows the plateaus for the expansion dynamics and the excellent agreement between theory and experiment. In the inset of Fig. \ref{expansion}(b) we also show density profiles for three successive measurement times inside a plateau. The entire density profiles almost perfectly overlap with each other, which clearly confirms that the expansion stops at the plateau.

For smaller $\lambda_z$, the trap frequency decreases slower, we find that the plateaus becomes denser and the difference in height between two adjacent plateaus becomes smaller. Eventually in the limit $\lambda_z\rightarrow 0$, one reaches the adiabatic limit and Eq. \ref{Rt_solution} reduces to $\mathcal{R}=\mathcal{R}_0 \sqrt{t/t_0}$. (See the supplementary material for discussions) For larger $\lambda_z$, the trap frequency drops faster and the plateau appears in a later time. In practice the experiment has a limit, that is, when the trap frequency is too below (say, below $30$Hz), it is  hard to get high quality data due to the low resolution of the images after the time-of-flight expansion. It prohibits us from observing plateaus for large $\lambda_z$ and investigating the critical regime $\lambda_z\rightarrow \lambda_\text{c}=4$ where the Efimovian expansion behavior disappears.

\begin{figure}[t]
\includegraphics[width=3.0 in]
{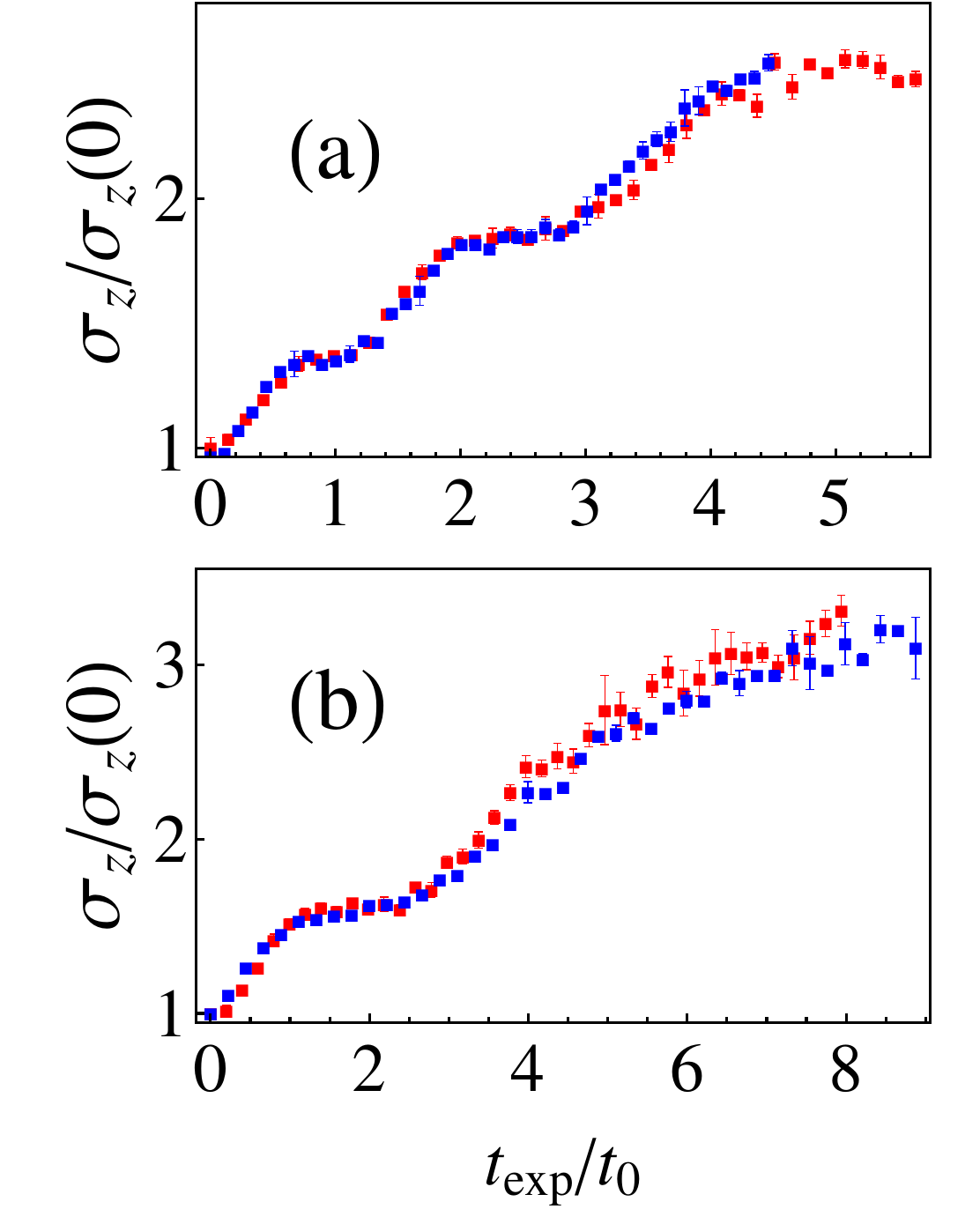}
\caption{$\sigma_z/\sigma_{z,0}$ as a function of $t_{\text{exp}}/t_0$ ($t_{\text{exp}}=t-t_0$) is universal for the non-interacting and the unitary Fermi gas, as long as they have the same $s_0$.  Blue dots and red dots are data measured for the unitary Fermi gas and the non-interacting Fermi gas, respectively. They have the same $s_0$ with $s_0=10.53$ in (a) and $s_0=5.88$ in (b). $\sigma_z$ is obtained from fitting the Gaussian density profile and $\sigma_{z,0}$ is the value of $\sigma_z$ at $t=t_0$. Error bars represent the standard deviation of the statistic.  }
\label{universal}
\end{figure}

Now we further demonstrate that this dynamics is universal. First, we should verify that $s_0$ relates to $\lambda_z$ via Eq. \ref{s0}. In the experiment,  $\lambda_z$ is determined by the trap frequencies measured by the parametric resonance and $s_0$ is extracted from the best fit of the expansion data in Fig. \ref{expansion}. The universal relation between $s_0$ and $\gamma\equiv\sqrt{1/\lambda_z-1/4}$ is plotted in Fig. \ref{s}(a).  $s_0(\gamma)$ can fit very well with a linear function $s_0=\kappa \gamma$ which gives the slope $\kappa=1.94\pm0.03$ for the non-interacting case and $\kappa=1.53\pm0.03$ for the unitary Fermi gas. These are in good agreement with $\omega_\text{b}=2$ for the non-interacting case and $\omega_\text{b}=\sqrt{12/5}=1.55$ for the unitary case. The Efimovian expansion is also robust and insensitive to the temperature and atoms number of the Fermi gas. Fig. \ref{s} (b) shows the dependence of  $s_0$ of the unitary Fermi gas on these parameters, where  $\lambda_z$ is fixed and both the temperature and the atom number are varied by controlling the evaporative cooling process. It shows that $s_0$ is almost a constant holding for the changing of the atoms number of each spin component from $5\times 10^4$ to $3\times 10^5$ and the dimensionless temperature $T/T_F$ varying from $0.61$ to $1.32$, where $T_F$ is the Fermi temperature.

\begin{figure}[t]
\includegraphics[width=3.0 in]
{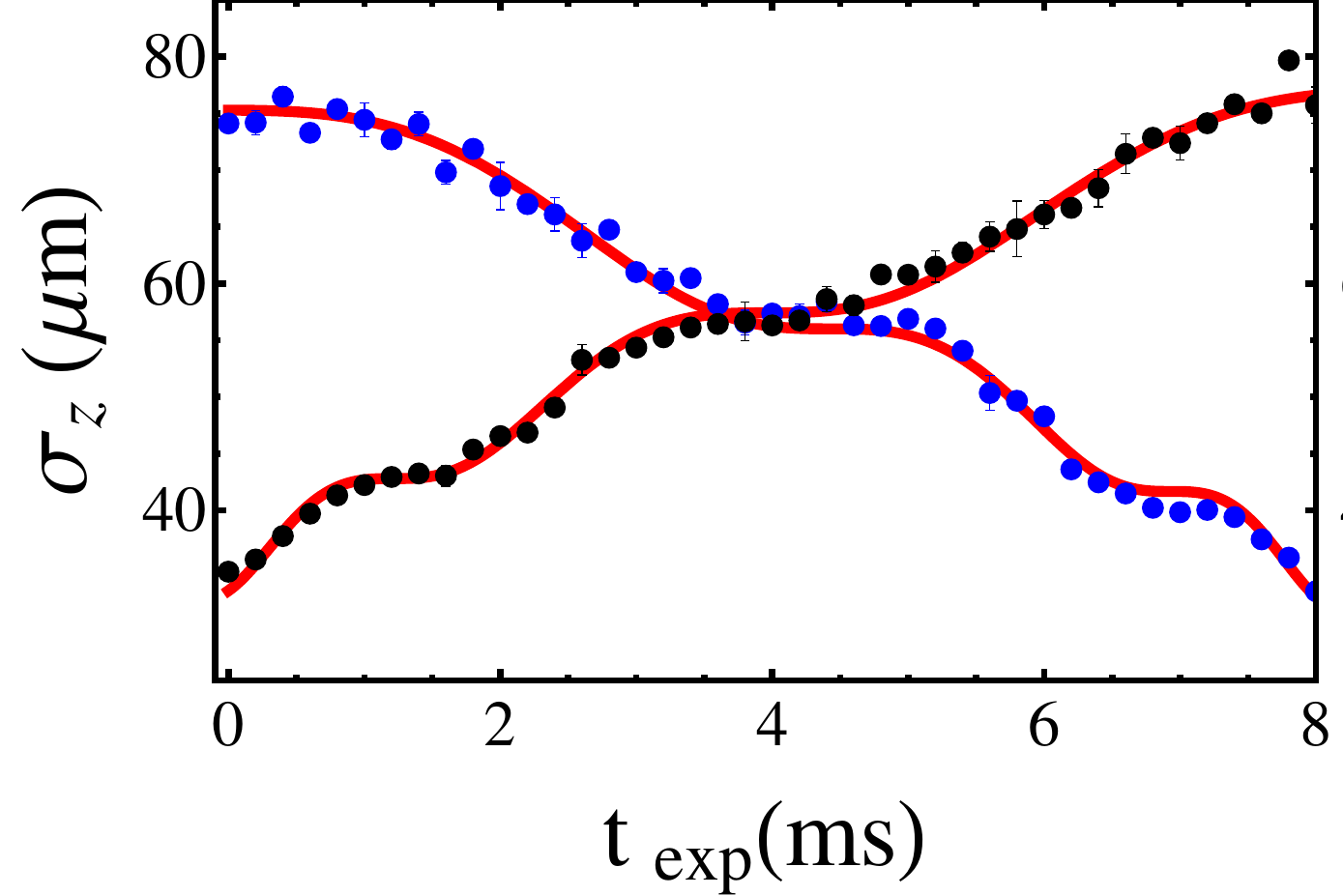}
\caption{$\sigma_z$  for the expansion and its inverted compression process from $t_0$ to $t_\text{f}$. $t_\text{exp}=t-t_0$. Black dots are the expansion process, with $\omega=1/(\sqrt{\lambda}t)$ and the frequency changing from $\omega_0=1/(\sqrt{\lambda}t_0)$ (at $t_0$) to $\omega_\text{f}=1/(\sqrt{\lambda}t_\text{f})$ (at $t_\text{f}$). Blue dots are the inverted compression process, with $\omega=1/(\sqrt{\lambda}(t_\text{f}+t_0-t))$ and the frequency changing from $\omega_\text{f}$ (at $t_0$) to $\omega_\text{0}$ (at $t_\text{f}$). Here $\lambda_z=0.01$ and the datas are taken in the unitary regime.  Error bars represent the standard deviation of the statistic.}
\label{time-invert}
\end{figure}

Second, we notice that the non-interacting and the unitary cases only differ in the relation between $s_0$ and $\lambda$, and once $s_0$ is given to be the same, the dynamics is the same for these two different systems. In another word, $\mathcal{R}_z/\mathcal{R}_z(0)$ as a function of $s_0$ (or $\varphi$) and $t/t_0$ is a universal function for all scale invariant state-of-matters. To better show this, we choose an expansion curve for the unitary Fermi gas and another expansion curve for the non-interacting gas with the same value of $s_0$, and we plot these two expansion curves together, as shown in Fig. \ref{universal}. For these two cases we have examined (two different $s_0$), the entire expansion curves perfectly coincide with each other. This proves experimentally that the Efimovian dynamics is universal for different scale invariant quantum gases.

At last, we study a time-reversal compression process. Consider an expansion process from $t_0$ to $t_\text{f}$, the trap frequency decreases from $\omega_0=1/(\sqrt{\lambda}t_0)$ to $\omega_\text{f}=1/(\sqrt{\lambda}t_\text{f})$. Now we consider an inverted process of increasing the trap frequency as $\omega=1/(\sqrt{\lambda}(t_\text{f}+t_0-t))$, the trap frequency increases from $\omega_\text{f}$ to $\omega_0$ when $t$ changes from $t_0$ to $t_f$. In order for the compression dynamics to really invert the expansion dynamics, $t_\text{f}$ has to be carefully chosen to satisfy $t_\text{f}=e^{2\pi n/s_0}t_0$. We perform such an experiment, and both the expansion and the compression dynamics are presented in Fig. \ref{time-invert} for comparison, which shows that the dynamical process with a carefully chosen boundary is time-reversal symmetric. The little asymmetry is because that the lowering trap (black dots) also plays a role as the evaporative cooling, which causes the atoms cooler and the cloud sizes smaller correspondingly.

In summary, we have discovered a novel type expansion dynamics for scale invariant quantum gases. It exhibits geometric scaling behavior, which is mathematically described by a log-periodic function, and displays plateaus features, which counter-intuitively shows that the cloud nearly stops expanding despite of the continuously decreasing of the confinement potential.  Moreover, it is \textit{universal} for all scale invariant quantum gases. Therefore, it can serve as a tool to detect the emergent scale invariance, similar as the anisotropic expansion now serving as a tool to probe the emergent hydrodynamics behavior. It can be used in future studies of dynamics in the quantum critical regime \cite{critical1,critical2,critical3,critical4,critical5} and to calibrate scale symmetry anomaly in a two-dimensional quantum gases \cite{anomaly1,anomaly2,anomaly3,anomaly4}.

\newpage
\widetext
\section{Supplemental Material}
\label{sec:supplement}

\subsection{The dynamical scaling solution}
\label{sec:hydro}
In the following, we consider a unitary Fermi gas trapped in a harmonic potential whose frequency is time dependent. The Hamiltonian is given by
\begin{eqnarray}
\hat{H}(t)=\hat{H}_0(t)+\hat{V}
\end{eqnarray}
where the non-interacting part
\begin{eqnarray}
\hat{H}_0(t)=\sum_{i=1}^{N}\bigg{[}-\frac{\nabla_i^2}{2}+\frac{x_i^2}{2\lambda_1t^2}+\frac{y_i^2}{2\lambda_2t^2}+\frac{z_i^2}{2\lambda_3t^2}\bigg{]}
\end{eqnarray}
represents the kinetic energy plus a time dependent harmonic trap, $N$ is total particle number and $\hat{V}=\sum_{{i\in\uparrow,{j\in\downarrow}}}V(\mathbf{r}_i-\mathbf{r}_j)$ represents the short range interaction between spin up and down fermions. Due to the divergence of the $s$-wave scattering length $a_\text{s}$, $\hat{V}$ is scale invariant in the zero interaction range limit such that $V(\Lambda r)=V(r)/\Lambda^2$. Thanks to this property of $\hat{V}$ and the specific choice of time dependence in $\hat{H}_0(t)$, the total Hamiltonian $\hat{H}$ has continuous scaling symmetry in both configurational and temporal spaces.

{\it Isotropic trapping.} To illustrate the basic idea and get a better physical intuition, we first consider the simpler isotropic trap with $\lambda_1=\lambda_2=\lambda_3=\lambda$. In this case, we can calculate the cloud size $\hat{R}^2=\sum_{i}\mathbf{r}_i^2/N$ directly from its equation of motion. The first derivative of $\langle \hat{R}^2\rangle$ is given as
\begin{eqnarray}
i\frac{d}{d t}\langle \hat{R}^2\rangle=\langle[\hat{R}^2,\hat{H}(t)]\rangle=\frac{2i}{N}\langle \hat{D}\rangle,
\end{eqnarray}
where $\hat{D}=\sum_{i}\frac{1}{2}(\mathbf{r_i\cdot p_i+p_i\cdot r_i})$ is the generator of a spacial scaling transformation. On the other hand, the equation of motion of $\langle\hat{D}\rangle$ is
\begin{eqnarray}
i\frac{d}{d t}\langle \hat{D}\rangle=\langle[\hat{D},\hat{H}]\rangle=2i\bigg{[}\langle \hat{H}(t)\rangle-N\frac{\langle \hat{R}^2\rangle}{\lambda t^2}\bigg{]}.
\end{eqnarray}
where we have used the fact that $V(r)$ is scale invariant such that $[\hat{D},\hat{V}]=2i\hat{V} $.
Combining these two equations we obtain
\begin{eqnarray}
\frac{d^2}{dt^2}\langle \hat{R}^2\rangle=4\left[\frac{1}{N}\langle \hat{H}(t)\rangle-\frac{\langle \hat{R}^2\rangle}{\lambda t^2}\right].\label{Rppt}
\end{eqnarray}
To make the equation closed, one still needs to calculate $d\langle \hat{H}(t)\rangle/dt$ which can be obtain by the Hellmann-Feynman theorem:
\begin{eqnarray}
\frac{d}{dt}\langle \hat{H}(t)\rangle=\left\langle\frac{d\hat{H}(t)}{dt}\right\rangle=-N\frac{\langle \hat{R}^2\rangle}{\lambda t^3}.\label{Hpt}
\end{eqnarray}
Combing (\ref{Rppt}) and (\ref{Hpt}), we finally obtain the following equation of motion for $\langle \hat{R}^2\rangle$
\begin{eqnarray}
\frac{d^3}{dt^3}\langle \hat{R}^2\rangle+\frac{4}{\lambda t^2}\frac{d}{dt}\langle \hat{R}^2\rangle-\frac{4}{\lambda t^3}\langle \hat{R}^2\rangle=0.\label{Rpppt}
\end{eqnarray}
Since this is a third order differential equation, one needs three initial conditions, which are the values of $\langle \hat{R}^2\rangle,~d\langle \hat{R}^2\rangle/dt$ and $d^2\langle \hat{R}^2\rangle/dt^2$ at the starting point $t=t_0$, to fix the solution. On one hand, since the system remains at static before the expansion, arbitrary order of time derivative of $\langle \hat{R}^2\rangle(t)$ remains 0 for all $t<t_0$. On the other hand, since $\hat{H}(t)$ is continuous while $d\hat{H}(t)/dt$ is discontinuous at $t=t_0$, it is easy to check that only the first and second derivative of $\langle \hat{R}^2\rangle(t)$ is continuous across $t_0$ while higher derivatives are discontinuous. Thus we will apply the following initial conditions
\begin{eqnarray}
\langle \hat{R}^2\rangle(t_0)&=&R_0^2\label{initial1}\\
\frac{d}{dt}\langle \hat{R}^2\rangle|_{t=t^+_0}&=&\frac{d^2}{dt^2}\langle \hat{R}^2\rangle|_{t=t^+_0}=0.\label{initial2}
\end{eqnarray}

The solution of (\ref{Rpppt}) has very different behavior for small and large value of $\lambda$. For $\lambda>\lambda_c=4$, we have
\begin{eqnarray}
\frac{\langle \hat{R}^2\rangle(t)}{R_0^2}=\frac{t}{t_0}\frac{\eta^2-1}{\eta^2}\left\{1-\frac{1}{2}\left[\frac{(t/t_0)^{\eta}}{\eta+1}-\frac{(t/t_0)^{-\eta}}{\eta-1} \right]\right\}
\end{eqnarray}
where $\eta=2\sqrt{1/4-1/\lambda}$. One can see that in the limit $t\gg t_0$, the cloud size follows a simple power law $\langle \hat{R}^2\rangle(t)\sim t^{1+\eta}$, where $\eta$ can be seen as an anomalous dimension in the time domain. As a result, the continuous scaling symmetry is still preserved for $\lambda>\lambda_c$ and there is no dynamic Efimov effect in this case.

The situation is much more interesting when $0<\lambda<\lambda_c$. In this case, we have
\begin{eqnarray}
\frac{\langle \hat{R}^2\rangle(t)}{R_0^2 }=\frac{t}{t_0\sin^2{\varphi}}\bigg{[}{1-\cos{\varphi}\cdot\cos\bigg(s_0\ln\frac{t}{t_0}+\varphi\bigg)}\bigg{]},\label{Rlogt}
\end{eqnarray}
where $s_0=2\sqrt{1/\lambda-1/4}$ and $\varphi=-\arctan{s_0}$. Instead of a simple power law form, $\langle \hat{R}^2\rangle(t)$ now contains a logarithmic periodic part which breaks the continuous scaling symmetry down to a discrete one in the time domain and satisfy
\begin{eqnarray}
\langle \hat{R}^2\rangle(te^{\frac{2n\pi}{s_0}})=e^{\frac{2n\pi}{s_0}}\langle \hat{R}^2\rangle(t)
\end{eqnarray}
for all integers of $n$.

{\it Adiabatic limit.} Now let us consider the adiabatic limit($\lambda\rightarrow0$) of the expansion. Physically, it represents the situation that the trap expands extremely slow. In this limit, Eq.~\ref{Rlogt} can be expressed as
\begin{eqnarray}
\frac{\langle \hat{R}^2\rangle(t)}{R_0^2 }=\frac{t}{t_0}\bigg{[}{1-\sqrt{\frac{\lambda}{4}}}\sin\bigg{(}s_0\ln\frac{t}{t_0}\bigg{)}+O(\lambda)\bigg{]}.
\end{eqnarray}
We find that the cloud size $\langle\hat{R}^2\rangle$ follows the size of the trap and grows proportional to $t$. This behavior is consistent with the adiabatic theorem which claims that the system remains in the instantaneous ground state of $\hat{H}(t)$.

{\it Anisotropic trapping.} Above approach, although elegant and rigorous, is only valid for an isotropic harmonic trap. For more general systems with anisotropic harmonic trap, there is no known exact solutions for the dynamic expansion. In this case, inspired by the logarithmic time dependence, we first perform the following scaling transformation
\begin{eqnarray}
t&=&t_0e^{\tau},\\
(x_i,y_i,z_i)&=&\sqrt{t}(u_i,v_i,w_i),\\
\psi(\{x_i,y_i,z_i\},t)&=&t^{-\frac{3N}{4}}\phi(\{\mathbf{q}_i\},\tau),
\end{eqnarray}
where $\mathbf{q}=(u,v,w)$ is the new position vector, and $N$ is the total number of fermions. This new set of space and time coordinates are now all dimensionless variables. After this change of variables, we obtain a new Hamiltonian for the transformed wave function $\phi(\{\mathbf{q}_i\},\tau)$
\begin{eqnarray}
i\partial_\tau\phi&=&\sum_i\bigg{[}\frac{(-i\nabla_i-\mathbf{q}_i/2)^2}{2}+\omega_1^2\frac{u_i^2}{2}+
\omega_2^2\frac{v_i^2}{2}+\omega_3^2\frac{w_i^2}{2}\bigg{]}\phi+\sum_{{i\in\uparrow,{j\in\downarrow}}}V(\mathbf{q}_i-\mathbf{q}_j)\phi. \label{Hstatic1}
\end{eqnarray}
Due to the scaling symmetry of the system, the original dynamic expansion problem is now transformed into a quantum quench problem under the new static Hamiltonian given by Eq.~\ref{Hstatic1}. This Hamiltonian corresponds to a unitary Fermi gas in a static harmonic trap with trapping frequency $\omega_k=\sqrt{1/\lambda_k-1/4}$ where $k=1,2,3$ and with a single particle gauge field $\mathbf{A}=\mathbf{q}/2$. The initial state $\phi(\{\mathbf{q}_i\},0)$ of this quench process is an eigen-state of another Hamiltonian
\begin{eqnarray}
\hat{h}&=&\sum_i\bigg{[}\frac{-\nabla_i^2}{2}+\frac{u_i^2}{2\lambda_1}+
\frac{v_i^2}{2\lambda_2}+\frac{w_i^2}{2\lambda_3}\bigg{]}+\sum_{{i\in\uparrow,{j\in\downarrow}}}V(\mathbf{q}_i-\mathbf{q}_j).
\end{eqnarray}
Since the gauge field $\mathbf{A}$ satisfies $\nabla\times\mathbf{A}=0$, it can be eliminated by a local gauge transformation $\phi(\{\mathbf{q}_i\},\tau)\rightarrow e^{i\sum_iq^2_i}\phi(\{\mathbf{q}_i\},\tau)$ which will introduce an initial current to $\phi(\{\mathbf{q}_i\},0)$ through the  phase factor $e^{i\sum_iq^2_i}$. When $\lambda_k\gg 1/4$, the quench will excite one or multiple breathing modes with small oscillation amplitude. In the axial symmetric case when $\omega_1=\omega_2=\omega_{\bot}$ and $\omega_3=\omega_z$, this problem can be solved with a hydrodynamic approach proposed in \cite{hydro}. In this method, the system is described by a set of density field such as the number density $n(\mathbf{q},\tau)$, entropy density $s(\mathbf{q},\tau)$. An exact solution to the two-fluid hydrodynamic equations was found in Ref. \cite{hydro} with a scaling ansatz
\begin{eqnarray}
n(\mathbf{q},\tau)=e^{2\alpha(\tau)+\gamma(\tau)}n_0(\mathbf{q'})
\end{eqnarray}
where $\mathbf{q'}=(e^{\alpha(\tau)}u,e^{\alpha(\tau)}v,e^{\gamma(\tau)}w)$ and the similar form of solution is also applied for $s(\mathbf{q},\tau)$. The axial size of the cloud for transformed wave function $\phi(\{\mathbf{q}_i\},\tau)$ is thus given as
\begin{eqnarray}
\langle q_z^2\rangle(\tau)=\int dq_xdq_ydq_z q_z^2 n(\mathbf{q},\tau)=e^{-2\gamma(\tau)}C
\end{eqnarray}
where $C=\int dq'_xdq'_ydq'_z (q'_z)^2 n_0(\mathbf{q'})$ is a constant. After transforming back to the original coordinates and when $\gamma(\tau)\ll 1$, one finally gets
\begin{eqnarray}
\langle \hat{R}_z^2\rangle(t)\simeq Ct\left[1-2\gamma\left(\ln\frac{t}{t_0} \right)\right].\label{Rt2}
\end{eqnarray}
As a result, the axial expansion of cloud is fully determined by the $\gamma(\tau)$ function.
 In the small perturbation regime, the functions $\alpha(\tau),~\gamma(\tau)$ satisfy the following linearized coupled differential equations \cite{hydro}
\begin{eqnarray}
\ddot{\alpha}+\omega_{\bot}^2\left(\frac{10}{3}\alpha+\frac{2}{3}\gamma \right)&=&0,\label{ab1}\\
\ddot{\gamma}+\omega_z^2\left(\frac{4}{3}\alpha+\frac{8}{3}\gamma \right)&=&0.\label{ab2}
\end{eqnarray}

It is obvious that the solution of $\gamma(\tau)$ takes a simple form $\gamma(\tau)=A_+\cos(\omega_{+}\tau+\phi_+)+A_-\cos(\omega_-\tau+\phi_-)$, with $A_\pm$ and $\phi_\pm$ to be fixed by initial conditions. Here $\omega_{\pm}$ is the eigen-frequency of (\ref{ab1}) and (\ref{ab2}) whose explicit expression was given in \cite{hydro}. In the limit of $\omega_{\bot}\gg\omega_z$, $\omega_{\pm}$ becomes $\omega_+=\sqrt{10/3}\omega_{\bot}$ and $\omega_-=\sqrt{12/5}\omega_z$. Moreover, the amplitude $A_+$ vanishes in this limit, thus we can subsitute this solution into (\ref{Rt2}) and implement the same initial conditions given in (\ref{initial1}) and (\ref{initial2}). We find that $\langle \hat{R}_z^2\rangle(t)$ are given by exactly the same expression as in (\ref{Rlogt}) but with a different $s_0=\sqrt{12/5}\omega_z=\sqrt{12/5}\sqrt{(1/\lambda_z-1/4)}$.

\subsection{Experimental methods}

Our experimental setup was presented in \cite{Wu}. A large atom number magneto-optical trap (MOT) is realized by a laser system of 2.5-watts laser output with Raman fiber amplifier and intracavity-frequency-doubler, as shown in Fig. 1(a).  After the loading stage, $5.2\times 10^9$ cold atoms are obtained.  Then MOT is compressed to obtain atoms of near Doppler-limited temperature of 280 $\mu K$ and a density of $10^{12}$ cm$^{-3}$. Following this compressed stage, the MOT gradient magnets are extinguished and repumping beams are switched off faster than the cooling beams. By optical pumping, a balanced mixture of atoms in the two lowest hyperfine states $|1\rangle\equiv|F=1/2, M=-1/2\rangle$ and $|2\rangle\equiv|F=1/2, M= 1/2\rangle$ is prepared.

The optical crossed dipole trap is consisted by a 200$\,$ W Ytterbium fiber laser at 1070$\,$nm. The resulting potential has a cylindrical symmetry around the propagation axis of the laser. The evaporative cooling is performed at the Feshbach resonance of the magnetic field $B=832$ Gauss.  The trap is turned on 100 ms before the MOT compressed stage. After the atoms are loaded into the dipole trap we hold the atoms 200 ms on the trap and then forced evaporative cooling is followed by lowering the trap laser power, as shown in Fig. 1(b). First a simple exponential ramp [$U_0\exp(-t/\tau)$] is used as a lowering curve, where $U_0$ is the full trap depth. The time constant $\tau$ is selected to control the trapping potential down to a variable value, which allows us to vary the final temperature and atom number of the cloud. After the forced evaporative cooling  the dipole trap is recompressed to a value $5\% \,U_0$. Then the trap depth is held 0.5 s for the equilibrium and is lowered  with the relation $U(t)\propto 1/t^2$. After the time $t$, the trap is turned off and a standard absorption image is followed to extract the cloud profiles. In the analysis of the images we use a Gaussian function to fit the column density and get the mean cloud size.

\begin{figure}
\begin{center}
\includegraphics[width=85mm]{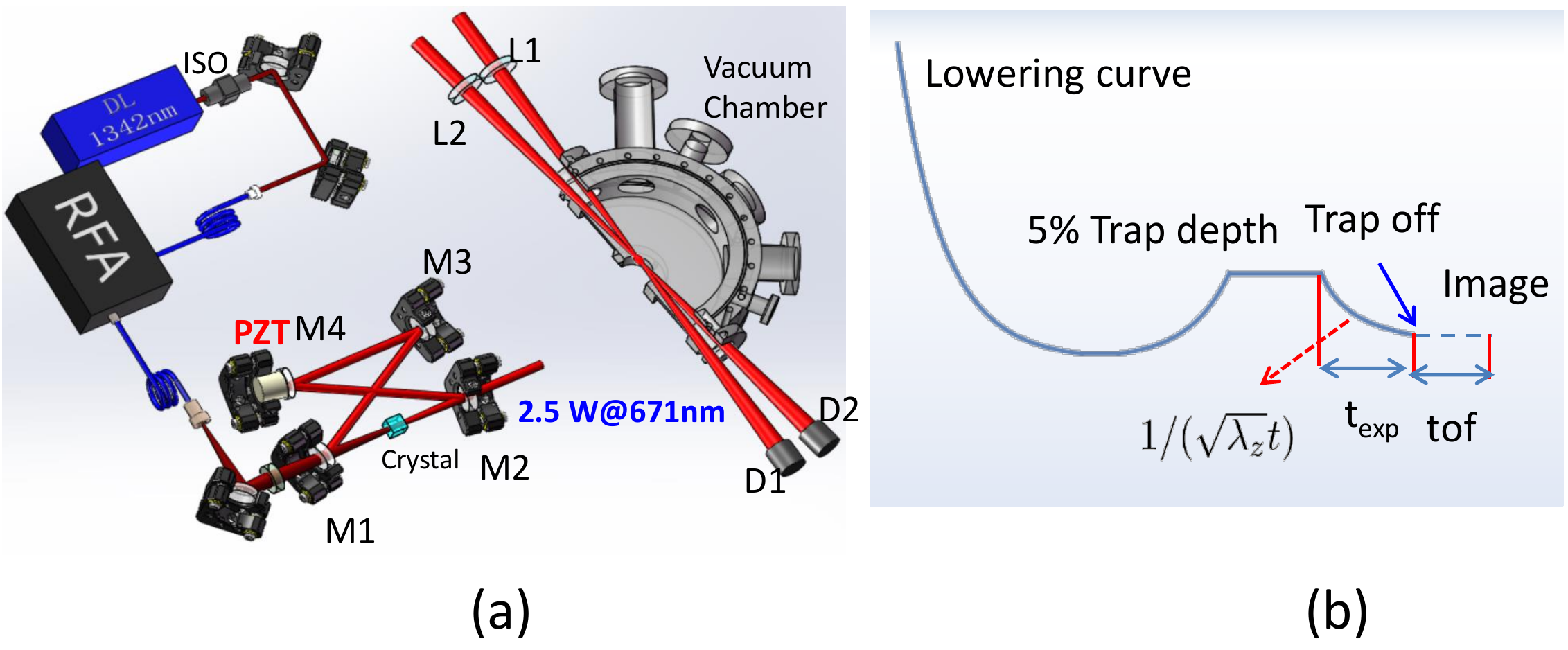}
\end{center}
\caption{The schematic of experimental setup (a) and the trap lowering curve (b).  RFA, Raman fiber amplifier; ISO, optical isolator; M1-M4, doubling-cavity mirrors; PZT, piezoelectric transducer; L1-L2, achromatic lenses; D1-D2, dumpers.
\label{fig:aspectratio}}
\end{figure}

\subsection{The expansion factor}
\label{sec:scaling}

The dynamical evolution is investigated by measuring the cloud radii after the cloud is released from the trap. So the size of the observed expanded cloud is related to that of the trapped gas by a scale factor $b_{\text{tof}}$, such that $\langle \sigma_z\rangle_{\text{obs}}=b_{\text{tof}}\langle \sigma_z\rangle$. The methods to analyze $b_{\text{tof}}$ of the non-interacting and the unitary Fermi gas are presented as follows:

{\it Non-interacting Fermi gas.} For the non-interacting Fermi gas, we can solve the dynamics of the time-of-flight exactly. Applying the Heisenberg equation to $\hat{R}^2_z$, we obtain
\begin{eqnarray}
\frac{d^2}{dt^2}\langle \hat{R}^2_z\rangle=4\langle \hat{T}_z\rangle.\label{tof1}
\end{eqnarray}
Here $\hat{T}_z=p_z^2/2m$ is the kinetic energy along the $z$-direction, which is a constant during the time-of-flight. We assume that the system is stationary before we release the trap. Then according to the Virial theorem,
\begin{eqnarray}
\langle\hat{T}_z\rangle=\langle V_z\rangle=\frac{1}{2}m\omega_{zf}^2\langle\hat{R}_{zf}^2\rangle,
\end{eqnarray}
where $\omega_{zf}$ and $\langle\hat{R}_{zf}^2\rangle$ is the trapping frequency and the cloud size just before the time-of-flight. Eq.~\ref{tof1} can be solved exactly and we obtain the scale factor
\begin{eqnarray}
b_{\text{tof}}=\sqrt{1+\omega_{zf}^2 t_{tof}^2},
\end{eqnarray}
where $t_{tof}=200\mu s$ is the flight time.

{\it Unitary Fermi gas.} To calculate the scale factor $b_{\text{tof}}$ of the unitary Fermi gas, we follow the hydrodynamic approach used in Ref. ~\cite{Johnscale}. First the density is given by
\begin{equation}
n({\mathbf{r}},t)=\frac{n_0(x/b_x,y/b_y,z/b_z)}{\Gamma},
\label{eq:density}
\end{equation}
where $b_i(t)$, $i=x,y,z$ is a time dependent scale factor. $n_0$ is the density profile of the trapped cloud before we start the Efimovian expansion. Here, $\Gamma(t)\equiv b_xb_yb_z$ is the volume scale factor, which is independent of the spatial coordinates in the scaling approximation. With Eq.~\ref{eq:density} and a velocity field that is linear in the spatial coordinates, $v_i=x_i\,\dot{b}_i/b_i$.
Here $\langle x_i^2\rangle=\langle x_i^2\rangle_0\,b_i^2(t)$, and $\langle v_i^2\rangle=\langle x_i^2\rangle\,\dot{b}_i^2/b_i^2=\langle x_i^2\rangle_0\,\dot{b}_i^2(t)$, where $\langle x_i^2\rangle_0$ is the mean-square cloud radius of the trapped cloud in the $i$-direction, just before releasing. Then, with these scaling assumptions and for the unitary Fermi gas, the fluids dynamics equations yield
\begin{equation}
\langle x_i^2\rangle_0\,b_i\,\ddot{b}_i=\frac{1}{Nm}\int d^3{\mathbf{r}}\,p-\frac{1}{m}\langle x_i \partial_i U_{\text{total}}\rangle,
\label{eq:3.1}
\end{equation}
where the heating from the shear viscosity is neglected. To add this term into the equation is straightforward, as shown in Ref.~\cite{Caoscience}.

\begin{figure}
\begin{center}
\includegraphics[width=85mm]{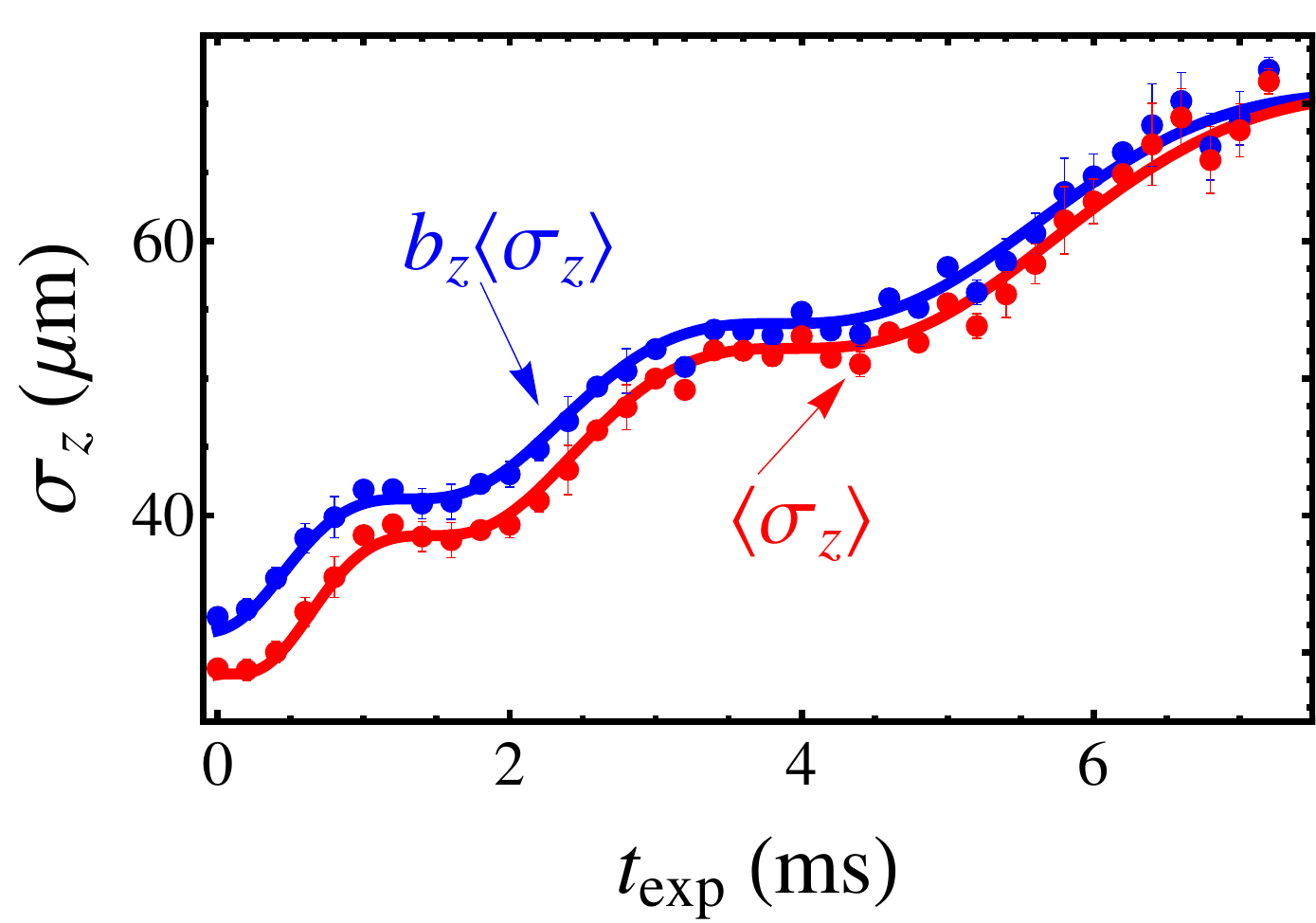}
\end{center}
\caption{$\langle \sigma_z\rangle$ obtained from the Gaussian fit of the density profile of the unitary Fermi gas for time-of-flight $t_{tof}=200\mu s$. $t_{\text{exp}}$ is the expansion time. The blue dots are the data $\langle \sigma_z\rangle_{\text{obs}}$ and red dots are $\langle \sigma_z\rangle=\langle \sigma_z\rangle_{\text{obs}}/b_{\text{tof}}$. The red solid line is the best fit based on theory curve. The blue solid line is the red solid line times by $b_{\text{tof}}$.  Error bars represent the standard deviation of the statistic.
\label{fig:expansionfactor}}
\end{figure}

Before the Efimovian expansion, the balance of the forces arising from the pressure and the trapping potential yields
\begin{equation}
\frac{1}{N}\int d^3{\mathbf{r}}\,p=\frac{\langle{\mathbf{r}}\cdot\nabla U_{\text{total}}\rangle_0}{3\,\Gamma^{2/3}}.
\label{eq:3.2}
\end{equation}
Substituting Eq.~\ref{eq:3.2} into Eq.~\ref{eq:3.1}, we get the evolution equations of the expansion factors,
\begin{equation}
\ddot{b}_i=\frac{\omega_{i0}^2}{\Gamma^{2/3}b_i}-\frac{\langle x_i \partial_i U_{\text{total}}\rangle}{m\langle x_i^2\rangle_0 b_i},
\label{eq:3.3}
\end{equation}
where
\begin{equation}
\omega_{i0}^2\equiv\frac{\langle x_i\partial_i U_{\text{total}}\rangle_0}{m\langle x_i^2\rangle_0}
=\frac{\langle{\mathbf{r}}\cdot\nabla U_{\text{total}}\rangle_0}{3m\langle x_i^2\rangle_0}.
\label{eq:3.4}
\end{equation}
Here, $U_{\text{total}}=U_{\text{opt}}+U_{\text{mag}}$ is the total trap potential including the optical trapping potential $U_{\text{opt}}$ and the residual magnetic potential $U_{\text{mag}}$ from the Feshbach coils. For the harmonic trap potentials, Eq.~\ref{eq:3.4} becomes
\begin{equation}
\ddot{b}_i=\frac{\omega_{i0}^2}{\Gamma^{2/3}b_i}-\omega_{\text{mag}}^2 b_i-\omega_{i,\text{opt}}^2(t) b_i,
\label{eq:3.5}
\end{equation}
where $\omega_{i0}^2$ is measured from the cloud profile and $\omega_{i,\text{opt}}$ is
\begin{eqnarray}
\omega_{i,\text{opt}}=\left\{
               \begin{array}{ll}
                 \frac{1}{\sqrt{\lambda_i}t}, & t_0\leq t\leq t_f \\
                 0, & t_f<t\leq t_f+t_{tof}.
               \end{array}
             \right.
\end{eqnarray}
Here $t_0$ is the time when we start the Efimovian expansion, $t_f$ is the time we turn off the trap and $t_{tof}$ is the time-of-flight. We numerically solve Eq.~\ref{eq:3.5} to determine $b_i$ with the initial conditions $b_i(t_0)=1$ and $\dot{b}_i(t_0)=0$. The scale factor of the time-of-flight is then $b_{\text{tof}}=b_z(t_f+t_{tof})/b_z(t_f)$. In Fig. \ref{fig:expansionfactor}, we show $\langle \sigma_z\rangle$ and $b_{\text{tof}}\langle\sigma_z\rangle$ of the unitary Fermi gas for flight time $t_{tof}=200\mu s$.

\end{document}